# AN UNIFORM DATA REPLICATION ALGORITHM IN WIRELESS MICRO-SENSOR NETWORK FOR COMMUNICATING MATERIALS APPLICATION

**Kais Mekki, William Derigent, Eric Rondeau**

Université de Lorraine, CRAN, UMR 7039
2 avenue de la forêt de Haye, Vandoeuvre-lès-Nancy Cedex, 54516, France
CNRS, CRAN, UMR 7039, France
kais.mekki8@etu.univ-lorraine.fr
william.derigent@univ-lorraine.fr
eric.rondeau@univ-lorraine.fr

**Ahmed Zouinkhi, Mohamed Naceur Abdelkrim**

National Engineering School of Gabes, Research Unit of Modeling, Analysis and Control of Systems (MACS)
rue Omar Ibn Elhkattab, 6029 Gabes, Tunisia
ahmed.zouinkhi@enig.rnu.tn
naceur.abdelkrim@enig.rnu.tn

**ABSTRACT:** *Wireless sensor networks (WSN) have recently gained a great deal of attention as a topic of research, with a wide range of applications being explored such as communicating materials. Data dissemination and storage are very important issues for sensor networks. The problem of designing data dissemination protocols for communicating material needs different analyses related to storage density and uniformity which has not been addressed sufficiently in the literature. This paper details storage protocol on the material by systematic dissemination through integrated wireless micro-sensors nodes. The performances of our solutions are evaluated through simulation using Castalia/OMNeT++. The results show that our algorithm provides uniform data storage in communicating material for different density level.*

## 1 INTRODUCTION

The communicating material is a new paradigm of industrial information system presented and discussed for the first time in (Kubler *et al.*, 2010). It enhances a classic material with the following capabilities: it can store data, communicate information at any point of its surface, and keep these previous properties after physical modifications. This concept leads to an important change in the internet of things. Indeed, the material does not communicate using some tags or nodes in specific points, but becomes intrinsically and continuously communicating. To meet this vision, many ultra-small electronic devices (thousands) are inserted into the material of the product during its industrial manufacturing.

The first works focusing on communication materials are presented in (Kubler *et al.*, 2013) (Kubler *et al.*, 2014). It introduces a communicating material (e-textile) obtained by scattering a huge amount of RFID *μtags* (1500 tags/m²) in a manufactured textile (i.e. *μtags* manufactured by the Hitachi company, measuring 0.15mmx0.15mm). The system involves a RFID reader/writer connected to a relational database that contains all the product life cycle information. At each writing operation, the database is explored to select the relevant data items (fragments of the database tables) that must be stored in the material. To do so, each data item is assigned an importance level between 0 and 1 (1=highly critical data item, 0=ordinary data item), computed via a multi-criteria decision-making algorithm. Data items with the highest importance levels are then stored in the *μtags* when the textile passes under the writer module during the manufacturing phases. Since the RFID are memory-constrained, the data item is splitted (segmented) and stored over several tags using a specific protocol header which is also able to rebuild the initial information. This division process is called segmentation and the resulting parts are segments.

With such system, the data storage in the communicating material requires a reader/writer connection with each tag. If a tag is not connected during the dissemination phase, it will be isolated and left empty which limits the use of RFID technology in solid and large materials such as concrete and wood. Therefore, our current works propose to use wireless sensor networks (WSN) in such products by spreading ultra-small and micro (even nano (Mahfuz and Ahmed, 2005)) sensor nodes into the material (e.g. 100 nodes/m²).

This paper presents dissemination algorithm to store relevant data items in wireless sensor nodes scattered into large scale products such as concrete in smart building (figure 1), plane wings or wood panels.

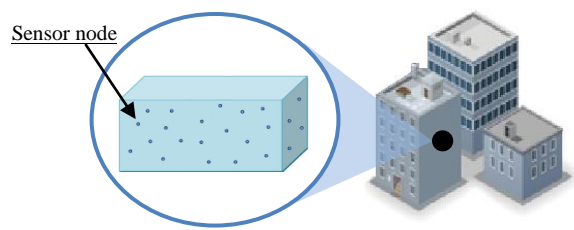

Figure 1: Application of communicating material concept in smart building

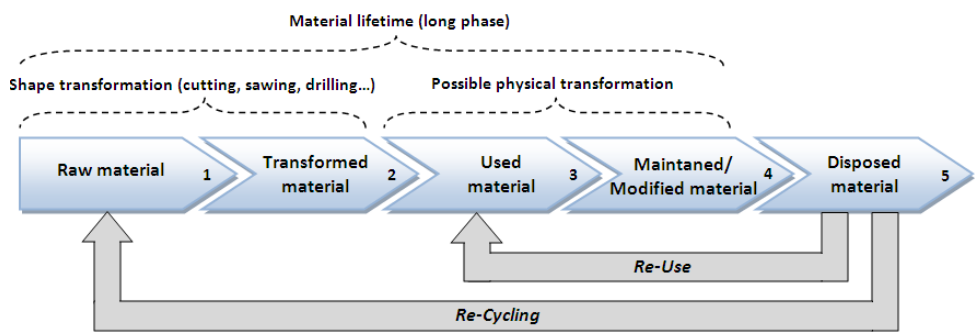

Figure 2: Communicating material life cycle

During the industrial construction of a final product and its future use, the communicating material has different life cycle steps as shown in figure 2. In steps 1 and 2, the material undergoes shape transformation (e.g. cutting, sawing, and drilling) to construct the product. Moreover, in the middle life cycle (steps 3 and 4), the material could have possible physical transformation (e.g. broken, piece addition/elimination…) which lead to information loss if data is not replicated in the whole communicating material. Therefore, the information should be stored into the material in uniform way (i.e. data is present in each piece of the material).

Indeed, in figure 3(a), the information is not uniformly disseminated, so pieces of the material are empty and information cannot be read after cutting as example. However, figure 3(b) shows a uniform replication where the information could be read in each piece.

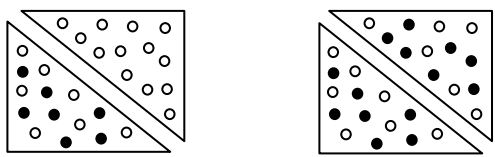

(a) Non-uniform dissemination    (b) Uniform dissemination

Figure 3: Uniform replication for anticipation of material transformation

As consequence, the dissemination algorithm must ensure that data is readable everywhere on the material even. A given data item should be uniformly replicated in the whole WSN of the material.

WSN in communicating material has some specificity:
- High node density (e.g. 100 nodes/m²),
- Energy-constrained nodes (Micro-nodes) with limited computation abilities,
- Nodes and batteries replacement is impossible (nodes embedded into the material).

Therefore, the dissemination algorithm in such networks and environment has to be judicious. The energy consumption has to be optimized by reduction of the amount of transmitted data in each node as much as possible (i.e. wireless transmission consumes more energy than any other communication activity (Ratnasamy *et al.*, 2002)). Hence, a long communicating material lifetime could be ensured during its life-cycle.

Another development imposition of the algorithm is the data importance level. To disseminate information, user sends data item coupled with an importance level. The number of storage nodes (i.e. storage density) should depend on this level as shown in figures 4(a) and 4(b). Therefore, the dissemination algorithm has to keep the uniformity of data item replication in the material whatever the importance level.

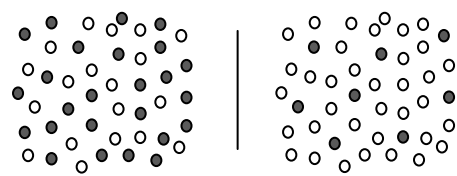

(a) Importance = 0.6    (b) Importance = 0.1

Figure 4: Storage density according to the importance level

The paper is organized as follow. Section 2 is dedicated to related works. Section 3 details the design of our algorithm for data dissemination and storage in communicating materials. Section 4 is dedicated to the presentation of simulation results and performance evaluation. Finally, section 5 concludes the paper.

## 2 RELATED WORK

In WSN, data dissemination protocols are used for network reprogramming, routing path discovery (e.g. DSR and AODV), fault tolerance, data mules, and mobile sink management (Akdere *et al.*, 2006). In past years, various algorithms have been proposed (Hongping and Kangling, 2010). There are two main approaches: reactive and proactive which is divided into structured and unstructured distribution strategies as shown in table 1. In the first one, nodes react to an event (e.g. the presence of mules and mobile sink) by disseminating data towards the nodes that are located close to the event positions, as example *Splash*, *CodeDrip*, *Typhoon*, *Supple*, and *Dynamic Random Replication* protocols. In the proactive approach, however, nodes distribute their data towards all or subset of nodes that have the role of storage unit, in anticipation of future event, node failure, mule collection, and others. Under structured dissemination, these storage nodes typically form a virtual structure (e.g. grid, line, and rail) within the WSN that make the data available to be retrieved later (e.g. by a mobile sink visiting the nodes to collect data), as example *LBDD*, *QDD*, *QAR*, *DDB*, *Locators*, and *ToW* protocols. However in unstruc-tured dissemination, the data is replicated throughout the whole network, as example *DEEP*, *RaWMS*, *RAPID*, *MDRW*, *MOAP*, *SSA*, *SDS*, and *McTorrent*.

|  | *Algorithms* | |
|---|---|---|
|  | ***Structured*** | ***Unstructured*** |
| ***Proactive*** | QDD (Mir and Ko, 2006), Locators (Shim and Park, 2006), LBDD (Hamida and Chelius, 2008), DDB (Lu and Valois, 2007), QAR (Rumin *et al.*, 2010), ToW (Joung and Huang, 2008)… | DEEP (Vecchio *et al.*, 2010), RaWMS (Bar-Yossef *et al.*, 2008), MDRW (Viana *et al.*, 2009), RAPID (Drabkin *et al.*, 2007), MOAP (Mukhtar *et al.*, 2009), McTorrent (Huang *et al.*, 2010), SSA (Chung *et al.*, 2011), SDS (Shen *et al.*, 2011)… |
| ***Reactive*** | Dynamic Random Replication (Cuevas *et al.*, 2010), Supple (Viana *et al.*, 2010), CodeDrip (Ribeiro *et al.*, 2014), Typhoon (Liang *et al.*, 2008), Splash (Doddavenkatappa *et al.*, 2013)… | |

Table 1: Taxonomy from a communication perspective

The aim of this work is to fully disseminate and to uniformly replicate data in the whole communicating material, in order to ease the reading process at any point of the material and to anticipate possible physical transformations. As a result, this section focuses on the protocols related to unstructured proactive dissemination.

An unstructured proactive dissemination protocols in WSN include a broadcast mechanism coupled with a storage strategy both are detailed in the following.

**2.1 Broadcast mechanisms**

Broadcast algorithms are usually referred to as flooding. Flooding is an important algorithm in WSN and is applied when a source node has to send information to subset or all nodes in the network. This is achieved by broadcasting a packet to the entire neighborhood. Each node that receives the packet rebroadcasts it, if it has not been forwarded previously. In this way, the information traverses the whole network and reaches all the nodes which then decide to store it or not according to the storage strategy. Although flooding is a very simple and efficient for data dissemination, it has some deficiencies. The main problems are: duplication (i.e. a node is forced to get information twice from two different nodes), collision (i.e. the broadcast increases the contention), and resource blindness (i.e. nodes do not adopt energy saving mechanisms) (Busnel *et al.*, 2007). As a consequence, various schemes for controlled flooding have been proposed including probabilistic-based also referred to as gossiping (Sekkas *et al.*, 2010) (Kyasanur *et al.*, 2006), counter-based (Miranda *et al.*, 2006) (Zhi-yan *et al.*, 2007), distance-based and location-based (Sanchez *et al.*, 2011).

Distance-based and location-based schemes exploit distance between nodes and their position. Therefore, nodes need to be equipped with a Received Signal Strength Indicator (RSSI) or a Global Positioning System (GPS). Such additional functions could not be applied in communicating materials since the nodes are embedded in the product with a high density (i.e. nodes are very close).

Authors in (Garbinato *et al.*, 2010) (Izumi *et al.*, 2007) show that the counter-based scheme outperforms the probabilistic one in terms of reliability (i.e. the percentage of nodes that are reached on average) and efficiency (i.e. the average amount of resources required for broadcasting a message). This mechanism can reduce the number of retransmitting nodes with a high arrival rate. Furthermore, it doesn't require specific hardware such as distance and location based schemes.

**2.2 Storage strategies**

Storage strategies in unstructured proactive dissemination protocols are developed in the literature according to the target application. In (Maatta *et al.*, 2010) (Dong *et al.*, 2012), the information is replicated in each node for network reprogramming. If a node does not receive the new version of binary code, its software is not updated and is isolated. Authors in (Maia *et al.*, 2013) (Neumann *et al.*, 2010) (Gonizzi *et al.*, 2013) propose storage strategies to improve network resilience against the risk of nodes failures. The storage nodes are selected according to some critical parameters such as connectivity, available memory and remaining energy. DEEP (Vecchio *et al.*, 2010) adopts another storing strategy for an effective data collection by mobile sink with uncontrolled trajectory. Every node that receives a data stores it with a probability *P* according to the number of information that must be replicated in each node. In (Bar-Yossef *et al.*, 2008), authors use a hop-counter mechanism to replicate data for mobile sink collection. Hop-counter is a process in which a data is repeatedly forwarded from a node to one of its chosen neighbors. At the same time, the number of hops realized from the source node is count. When the counter is equal to a predefined value, the information is stored in the current node.

**2.3 Synthesis**

The above storing strategies use the physical nodes parameters (e.g. memory and energy), network topology, and neighborhood awareness to replicate data throughout the WSN. They don't assume the properties and the characteristic of the disseminated data itself. This problem is the key contribution of our work, in which a dynamic information-based dissemination protocol is proposed and applied on the communicating material framework.

In this current paper, the data importance level is processed. The item with a high level must be replicated more than the lower one. The dissemination algorithm has to ensure the uniformity of replication for each level. For these reasons, probabilistic approach coupled with counter-based flooding is developed in this article and the results are evaluated from a uniformity and energy consumption point-of-view.

The developed solution represents an integration of a probabilistic storage approach in an energy-efficient counter-based broadcasting scheme. The replication process is controlled within each neighborhood using a packet header named Node-Storage (*NS*). All data seg-ments, hence the information, are forced to be stored only once in every neighborhood which ensure the same storage density and highly uniformity throughout the communicating material.

The algorithm is simulated with Castalia/OMNeT++ using a realistic collision model. It is evaluated by studying the uniformity of replication for different importance levels and the energy consumption rate.

The rest of the paper details the proposed data dissemination and replication algorithm, and then presents simulation results obtained for different data item importance level.

## 3 DATA REPLICATION ALGORITHM

This section presents our algorithm of data item dissemination. Firstly, the concept of master node is defined, then the counter-based forwarding scheme is described, and finally the dissemination/replication algorithm is presented.

### 3.1 Master Node

When the user has to disseminate information, he connects to the communicating material through a selected node in its transmission range which is named in this paper "master node" (i.e. *any node in the material could be a master node*). The Master node is responsible for forwarding user requests, collecting response messages from all nodes, and disseminating data items. Some constraints could be used to select the master node such as the highest residual energy level.

### 3.2 Counter-based scheme

In simple flooding, the more duplicate broadcasted packets a node receives, the less effective its rebroadcasting becomes (Lu and Valois, 2007). This is because the duplicate messages are likely to have been received by its neighboring nodes. This fact is involved in the counter-based flooding. In this scheme, a node that has received redundant packet more than a predefined threshold $C_{th}$ cancels the rebroadcasting. The counter-based algorithm is shown below:

1. When a node receives a broadcasted packet for the first time, the node initializes a counter $N$ to one, and sets a random assessment delay (*RAD*) at a value between 0 and $T_{max}$.

2. If the node receives the same broadcasted packet during the *RAD*, it increments the counter $N$ ($N\leftarrow N+1$). If the counter reaches a preset threshold $C_{th}$, it cancels the rebroadcasting.

3. After the *RAD* expires, the node rebroadcasts the packet to all neighbors.

Figure 5 summarizes the algorithm of this scheme in each node.

| *Algorithm: Counter-Based Broadcasting Scheme* |
|---|
| For a node *X* |
| Upon reception of a broadcasted packet *m* for the first time |
|   - Initialize the packet counter *N* to 1 |
|   - Set and wait for *RAD* to expire |
|   - While waiting: |
|     • For every duplicate packet *m* received |
|     • Increment *N* by 1 |
|   - When the delay RAD expires |
|     - If (*N<Cth*) |
|       • Retransmit the packet *m* |
|     - Else |
|       • Drop the packet *m* |
| End Algorithm |

Figure 5: Algorithm of the counter-based broadcasting scheme

This scheme is improved in (Mekki *et al.*, 2014) for a long communicating material lifetime. The random waiting delay is processed to prevent as much as possible the nodes with low energy level from rebroadcasting. As consequence, the rate of rebroadcasting of each node is adapted to their remaining energy. In the original counter-based scheme, node selects random waiting delay *RAD* between 0 and a fixed period of time $T_{max}$. To adopt this mechanism to the remaining energy level of node, the equation 1 and 2 are developed.

$$RAD_{extension} = RAD \times K \quad (1)$$

$$K = (E_{initial} / E_{remainder}) \quad (2)$$

$RAD_{extension}$: random assessment delay extension.
$E_{initial}$: initial energy of sensor node.
$E_{remainder}$: remainder energy of sensor node.

*RAD* is multiplied by *K* which increases the waiting delay when the remainder energy becomes lower as shown in figures 6 and 7.

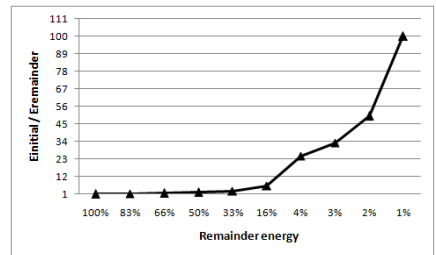

Figure 6: Remainder energy versus the fraction $E_{initial}/E_{remainder}$ for $E_{initial}$=100J

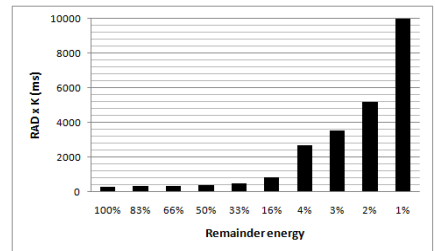

Figure 7: Remainder energy versus *RAD*x*K* for *RAD*=100ms

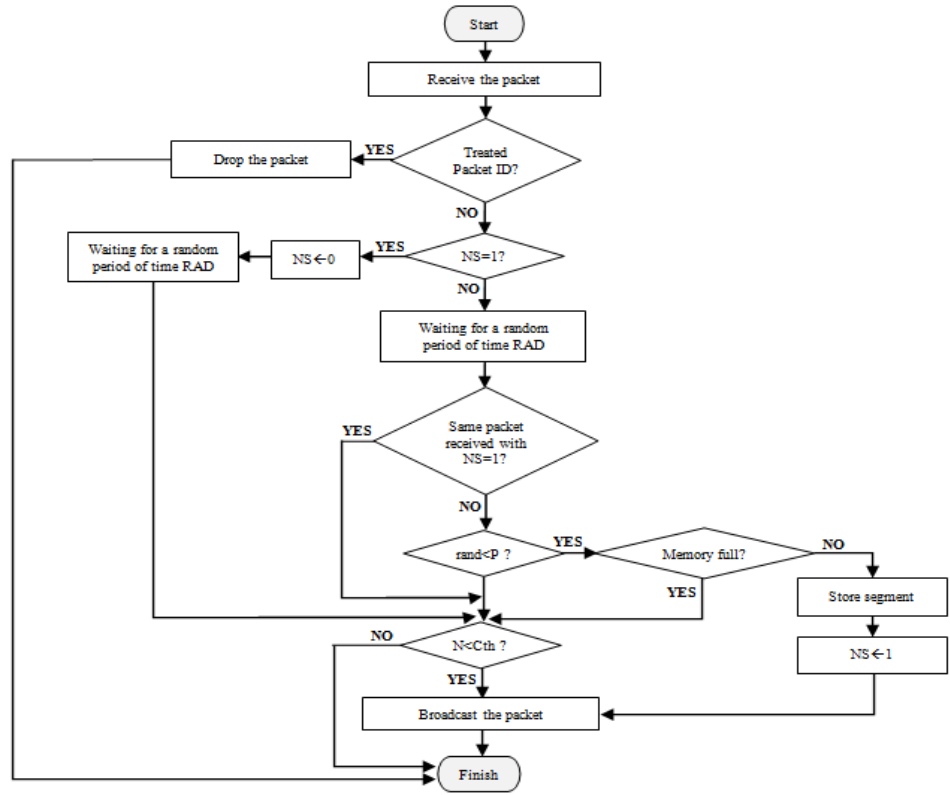

Figure 8: Data replication algorithm in communicating material

If the remainder energy is very low, the waiting delay $RAD_{extension}$ is larger. So, the node has to wait more than others and count the number of retransmission $N$ of neighbor nodes which have more energy. As consequence, it has more chance to reach the counter threshold $Cth$ and cancel the rebroadcast.

Simulation results in (Mekki *et al.*, 2014) showed that this delay extension maintain the high reachability of the original scheme. And, it enhances the communicating material lifetime by reducing the number of retransmitting nodes by about 10% (i.e. optimizes more the energy consumption than the original counter-based scheme). Moreover, the optimal parameters of the improved scheme was fixed to $T_{max}$=200ms and $Cth$=4 for high reachability and number of retransmitting node trade-off. Such parameters are consistent with previous optimized interval values for the counter-based scheme as described in (Arango *et al.*, 2006) and (Jacobsson *et al.*, 2011).

### 3.3 Storage strategy

The master node starts the dissemination process. The packet is broadcasted from one node to all its neighbors using the improved energy-efficient counter-based flooding scheme. The storage strategy is integrated in this broadcasting scheme. The node stores the data with a probability $P$ which is equal to the importance level as shown in figure 9 and equation 3 (i.e. the master the node start dissemination process by associating the probability $P=I$ to the data).

$$F : [0,1] \rightarrow [0,1]$$
$$I \rightarrow P=I \quad (3)$$

With $I$ is the importance level of the data item.

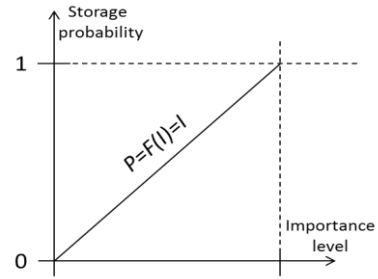

Figure 9: Conversion function of data importance to storage probability

To ensure a uniform and controlled replication rate of data segments, a packet header named Node-Storage (*NS*) is defined which takes two values 0 and 1. When a node stores a data, it rebroadcast packet with *NS*=1 to inform neighbor nodes, hence replicate data once within each neighborhood. Else, *NS* is set to 0.

The extended scheme works as follows:

1. When a node receives a broadcasted packet for the first time, the node initializes a counter $N$ to one, and sets a *RAD* at a value between 0 and $T_{max}$. If the received packet has *NS*=1, the storage decision is cancelled and *NS* is set to 0.

2. During the *RAD*:
   - If the node receives the same broadcasted packet with *NS*=1, the storage decision is cancelled and *NS* is set to 0.
   - If the counter reaches the preset threshold $Cth$, it cancels the rebroadcasting.

3. After the *RAD* expires:

- The node stores the data with a probability *P*, computed with the conversion function *F* from the importance level (see table 2), put *NS* to 1 and rebroadcast the packet even if the counter *Cth* of broadcasting scheme is reached. As an example, let *P*=0.5. The node selects a random value *rand* within the interval [0,1]. The node stores the data item only if *rand≤P*(=0.5).
- If the storage decision is cancelled and the counter *Cth* is not reached, the node rebroadcast the packet to all neighbors.

This process continues until the edges of material are reached. Our dissemination algorithm is broadcast based. With such communication model, the packet is further processed by nodes that have already received it. Furthermore, user can disseminate many segments or other data items. So, each packet is identified in the network layer (Packet *ID*) and then the nodes are limited to accept the same message only once. If a node receives again a treated message with the same *ID*, it will be dropped. Figure 8 shows the flowchart of this process executed in each node.

## 4 SIMULATION AND PERFORMANCE EVALUATION

This section describes the details of the simulation setup, the used parameters and finally simulation results of data replication.

### 4.1 Simulation setup

The performance of our solution is evaluated through simulation. Castalia/OMNET++ simulator is used to implement our dissemination algorithms. Currently, many wireless sensor network simulators are available as COOJA and TOSSIM but Castalia provides realistic wireless channel, radio models, and node behavior (Sundani *et al.*, 2011).
We simulate the Tyndall 10mm sensor node (Shen *et al.*, 2009) which is one of the smallest nodes used in the industry and research (10mm x 10mm). Thus, a thousand of scattered nodes in small material size could be simulated. Table 2 and figure 10 show the characteristic of Tyndall 10mm node.

| Size | 10 mm by 10 mm | |
|---|---|---|
| Microcontroller | Atmega128 | Single chip transceiver/microcontroller |
| Transceiver | Nordic nRF905 | |
| Frequency bands | 433 MHz, 868 MHz, and 915 MHz | |
| Bandwidth | 50 kbps | |
| Memory | 4 kb | |
| Microcontroller Unit | 688 µA | |
| Transmission mode | 10.35 mA | |
| Reception mode | 13.32 mA | |
| Moderate sleep | 166 µA | |
| Deep sleep | 3.3 µA | |

Table 2: Tyndall 10mm node characteristics

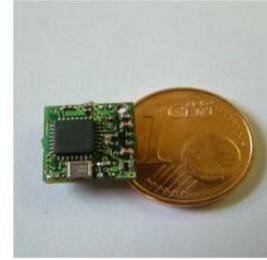

Figure 10: Tyndall node

The simulated communicating material consists of 2500 nodes. The nodes positions are all uniformly distributed within a 50mx50mx1m square (1 node/m$^3$) and the transmission power -25 dbm is used which ensures that a given node can communicate with 25 neighbors in average. The chosen simulation parameters are summarized in table 3.

| Material size | 50mx50mx1m | |
|---|---|---|
| Number of nodes | 2500 | |
| Nodes distribution | Grid 50x50 | |
| Nodes density | 1 node/m$^3$ | |
| Average neighborhood density | 25 nodes | |
| Location of master node | Center of the material | |
| Mac Protocol | TMAC | |
| Sigma channel parameter | 4 | Real radio wireless channel in Castalia |
| BidirectionalSigma channel parameter | 1 | |
| Radio Collision Model | Additive interference model | |
| Cth | 4 | |
| T$_{max}$ | 200 ms | |
| Number of data segments | 3 | |
| Simulation time | 100s | |
| Number of trials | 60 | |

Table 3: Simulation parameters

### 4.2 Simulation results

The impacts of the storage strategies on the uniformity performance for the proposed data dissemination algorithms have been investigated, through Castalia simulation. Figure 11 and 12 show the simulation results of the broadcast and storage mechanisms of segmented data item for different importance level (i.e. different storage probability).
On the basis of the obtained results, the following observation can be carried out. The data is well replicated in the communicating material according to the importance level (probability of storage). When a low importance is used, as example *P*=0.1, data item is stored in a maximum of 380 nodes in the communicating material. Furthermore, the number of storage nodes does not highly increased for large probability value, Moreover, segments are replicated with the same density (no segment is stored much than others) as a controlled neighborhood storage strategy is used (i.e. the segment is stored once in each neighborhood as much as possible).

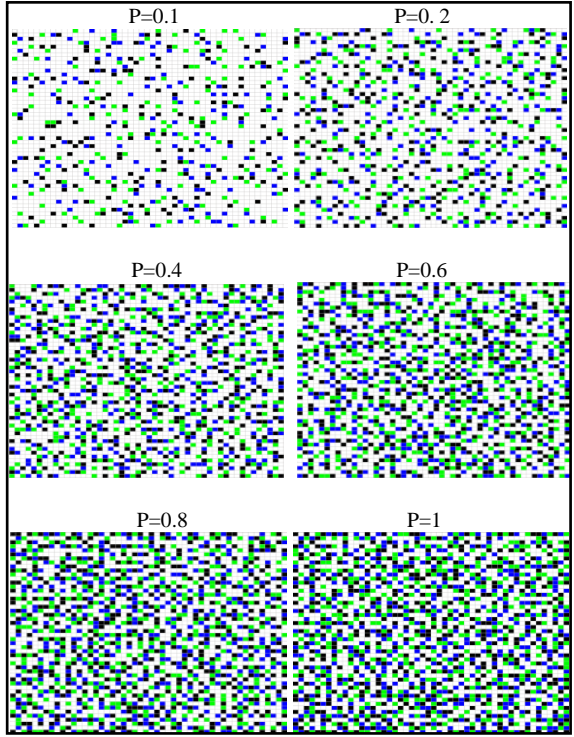

Figure 11: Simulation results

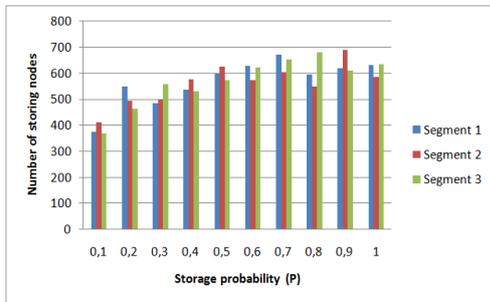

Figure 12: Storage density of each segment for different probability

**4.3 Uniformity and energy performances**

The uniformity and energy consumption performance of our algorithm is compared to DEEP which is a recent protocol of data dissemination and replication in WSN.
DEEP strategy is based on efficient probabilistic forwarding and storing strategy. The probabilistic forwarding is inherited from the RAPID dissemination mechanism (Drabkin *et al.*, 2007). In particular, the data dissemination strategy of this approach employs a combination of density sensitive probabilistic forwarding with deterministic corrective measures, as described in (Drabkin *et al.*, 2007). Essentially, the goal of RAPID is to ensure that there will be a predefined average number of retransmissions of each message in each neighborhood. This is obtained as follows: each node $p$ that receives a message $m$ for the first time, decides to rebroadcast $m$ immediately with a probability $F$ presented in equation 4:

$$F = \min\left(1, \frac{\beta}{|N(p)|}\right) \quad (4)$$

Where $N(p)$ is the one-hop neighborhood of $p$ (i.e. the number of nodes within the radio transmission range of $p$) and $\beta$ is the desired average number of retransmissions in each neighborhood. Additionally, to overcome situations in which, due to the probabilistic nature of this process, no node decides to transmit in a given neighborhood, RAPID complements this with a semi-deterministic corrective measure. Specifically, if a given node $q$ decides not to retransmit $m$ but does not hear any other retransmission of $m$ after a (relatively long) randomly selected period of time, then $q$ rebroadcast $m$ after all.

So, DEEP take the above described broadcasting scheme of RAPID and couple it with a storing strategy by which every node that receives a message carrying the disseminated data, stores it with probability $P$.

DEEP is implemented in Castalia, and is simulated for each storage probability $P$ to replicate segments in the whole communicating material.

*4.3.1 Uniformity performance*

To study the uniformity performance, a neighborhood window is used as shown in figure 13.

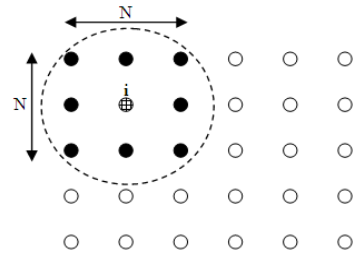

Figure 13: Uniformity evaluation method

The N×N window represents the one-hop neighborhood of a node $i$. The window is moved around all nodes in the material, and each time, we check the existence of segments within this window (i.e. the information exists in each neighborhood if the three segments are present).
The 5×5 window is the window that analyzes the performance of our simulation scenario as the simulated average neighborhood density is equal to 25 nodes.
Figures 14, 15, and 16 show the uniformity performance for windows 5×5 (25 nodes), of our proposed algorithm and DEEP protocol.

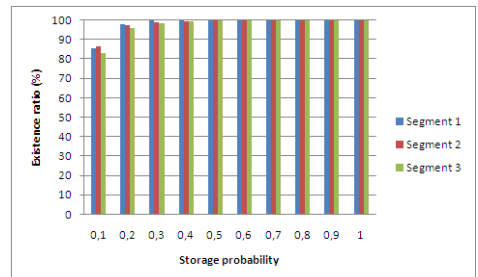

Figure 14: Existence ratio of segments in each neighborhood for the proposed algorithm

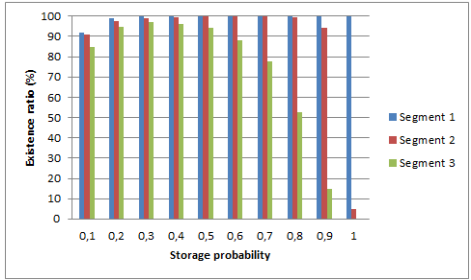

Figure 15: Existence ratio of segments in each neighborhood for DEEP

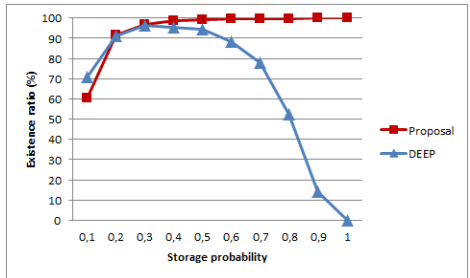

Figure 16: Uniformity performance of the proposed algorithm versus DEEP

As shown in figure 14 and 16, all segments are presents and hence the information exists for 100% of nodes in the material for a probability higher than 0.4, and, above 93% for storage probability between 0.2 and 0.3. However, the uniformity decrease for 0.1 as the used storage probability is very low (i.e. the information exists for 60% of nodes).

For DEEP algorithm, if the storage probability is increased above 0.4, the uniformity decrease, as the segments are not replicated with the same density as shown in figure 15. The first disseminated segment is replicated more than others, for example the segment 3 is not stored in any nodes for probability $P=1$ as all nodes are occupied by first segments, hence the information could not be read in any part of the material (information existence ratio=0%).

This shows the advantage to use NS packet header in our algorithm because it allows controlling the segment storage rate in the material. All segments are replicated for any storage probability within each neighborhood.

So, the proposed algorithm is considered as highly uniform and the information could be read in each pieces of the material even after a shape transformation. Moreover, for potential future reading (step 3 of the material life cycle (figure 2)), the master node won't need to send request for more than 1 hop or 2 hops in the material (i.e. the multi-hop connectivity is not required for stored data gathering). The information could be read using a simple neighborhood request/response, which optimize and reduce the energy consumption as no long communication is needed (i.e. energy efficient information retrieval).

### 4.3.2 Energy consumption performance

In this study, the average consumed energy by all nodes is measured after the end of the dissemination/replication process. The consumed energies of the proposal algorithm and DEEP are shown in figure 17.

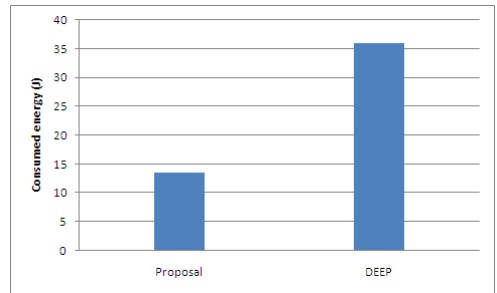

Figure 17: Energy consumption performance of the proposed algorithm versus DEEP

Figure 17 shows that our algorithm consume less energy than DEEP after the dissemination process, as an energy-efficient counter-based scheme is used to broadcast packets in the whole material. The probabilistic-based forwarding scheme of DEEP is not a deterministic scheme which could lead to many collision and interference, hence retransmissions and more energy consumption. The random delay extension adopted for our algorithm keeps a lot of energy and gives an equitable use of nodes resources (energy balance).

Therefore, our solution optimizes more the energy consumption and ensures a long communicating material lifetime (i.e. network lifetime) for a dissemination and replication process.

## 5 CONCLUSION

In this paper, a data dissemination algorithm is proposed to replicate information in communicating materials through scattered ultra-small wireless sensor nodes. A counter-based flooding approach and probabilistic strategy are used to distribute the packets to the whole network and to control the replication rate according to the importance level of the user information. The performances of the solution are evaluated through computer simulation. Simulation results show that our algorithm provides uniform replication for different storage probability. However, if a very low importance value is used (0.1) on the dissemination phase, large empty areas and low uniformity are obtained which can lead to data loss during physical material transformation. Our future works will be focused on testing the proposed algorithm in real WSN using IoT-LAB platform (www.iot-lab.info).

## REFERENCES

Akdere, M., C.C. Bilgin, O. Gerdaneri, I. Korpeoglu, O. Ulusoy, and U. Cetintemel, 2006. A comparison of epidemic algorithms in wireless sensor networks. *Journal of Computer Communications*, 29, p. 2450-2457.

Arango, J., A. Efrat, S. Ramasubramanian, M. Krunz, and S. Pink, 2006. Retransmission and Back-off Strategies for Broadcasting in Multi-hop Wireless

Networks. *3rd IEEE International Conference on Broadband Communication, Networks and Systems*, San José, California, USA.

Bar-Yossef, Z., R. Friedman, and G. Kliot, 2008. RaWMS: RandomWalk based Lightweight Membership Service for Wireless Ad Hoc Networks. *ACM Transactions on Computer Systems*, 26(2), p. 1-66.

Busnel, Y., A. Ghodsi, K. Iwanicki, H. Miranda, and H. Weatherspoon, 2007. Gossiping over storage systems is practical. *ACM SIGOPS Operating Systems Review*, 41(5), p. 75-81.

Chung, Y.-C., I.F. Su, and C. Lee, 2011. An efficient mechanism for processing similarity search queries in sensor networks. *Information Sciences Journal*, 181(2), p. 284–307.

Cuevas, A., M. Uruena, and G. de Veciana, 2010. Dynamic Random Replication for Data Centric Storage. *13th ACM international Conference on Modeling, Analysis, and Simulation of Wireless and Mobile Systems*, Bodrum, Turkey, p. 393-402.

Doddavenkatappa, M., M. C. Chan, and B. Leong, 2013. Splash: Fast Data Dissemination with Constructive Interference in Wireless Sensor Networks. *10th USENIX conference on Networked Systems Design and Implementation*, Lombard, USA, p. 269-282.

Dong, W., C. Chen, X. Liu, G. Teng, J. Bu, and Y. Liu, 2012. Bulk data dissemination in wireless sensor networks: Modeling and analysis. *Computer Networks Journal*, 56, p. 2664-2676.

Drabkin, V., R. Friedman, G. Kliot, and M. Segal, 2007. RAPID: reliable probabilistic dissemination in wireless ad hoc networks. *26th IEEE International Symposium on Reliable Distributed Systems (SRDS)*, Beijing, China, p. 13-22.

Garbinato, B., D. Rochat, M. Tomassini, and F. Vessaz, 2010. Injecting power-awareness into epidemic information dissemination in sensor networks. *Journal of Future Generation Computer Systems*, 26, p. 868-876.

Gonizzi, P., G. Ferrari, V. Gay, and J. Leguay, 2013. Data dissemination scheme for distributed storage for IoT observation systems at large scale. *Journal of Information Fusion*.

Hamida, E.B. and G. Chelius, 2008. A line-based data dissemination protocol for wireless sensor networks with mobile sink. *IEEE International Conference on Communications (ICC 2008)*, Beijing, China.

Hongping, F. and F. Kangling, 2010. Overview of data dissemination strategy in wireless sensor networks. *International Conference on E-Health Networking, Digital Ecosystems and Technologies*, Shenzhen, China, p. 260-263.

Huang, L., S. Setia, and R. Simon, 2010. McTorrent: Using multiple communication channels for efficient bulk data dissemination in wireless sensor networks. *Journal of Systems and Software*, 83, p. 108-120.

Izumi, S., T. Matsuda, H. Kawaguchi, C. Ohta, and M. Yoshimoto, 2007. Improvement of Counter-based Broadcasting by Random Assessment Delay Extension for Wireless Sensor Networks. *IEEE International Conference on Sensor Technologies and Applications*, Valencia, Spain, p. 76-81.

Jacobsson M., C. Guo, and I. Niemegeers, 2011. An experimental investigation of optimized flooding protocols using a wireless sensor network testbed. *Journal of Computer Networks*, 55, p. 2899-2913.

Joung, Y.-J. and S.-H. Huang, 2008. Tug-of-War: An Adaptive and Cost-Optimal Data Storage and Query Mechanism in Wireless Sensor Networks. *Journal of Distributed Computing in Sensor Systems*, 5067, p. 237-251.

Kubler, S., W. Derigent, A. Thomas, and E. Rondeau, 2010. Problem definition methodology for the Communicating Material paradigm. 10th IFAC Workshop on Intelligent Manufacturing Systems, Lisbonne, Portugal.

Kubler, S., W. Derigent, A. Thomas, and E. Rondeau, 2013. Embedding data on "communicating materials" from context-sensitive information analysis. *Journal of Intelligent Manufacturing*.

Kubler, S., W. Derigent, A. Voisin, A. Thomas, and E. Rondeau, 2014. Method for embedding context-sensitive information on "communicating textiles" via fuzzy AHP. *Journal of Intelligent and Fuzzy Systems*, 26(2), p. 597-610.

Kyasanur, P., R. Choudhury, and I. Gupta, 2006. Smart gossip: an adaptive gossip-based broadcasting service for sensor networks. *IEEE International Conference on Mobile Ad Hoc and Sensor Systems*, Vancouver, British Columbia, Canada, p. 91-100.

Liang, C.-J., R.-E. Musaloiu, and A. Terzis, 2008. Typhoon: A Reliable Data Dissemination Protocol for Wireless Sensor Networks. *Computer Science Journal*, p. 268-285.

Lu, J.L. and F. Valois, 2007. On the data dissemination in wsns. *3rd International Conference on Wireless*

and Mobile Computing, Networking and Communications (WiMob)*, New York, USA.

Maatta, L., J. Suhonen, T. Laukkarinen, T.D. Hamalainen, and M. Hannikainen, 2010. Program image dissemination protocol for low-energy multihop wireless sensor networks. *International Symposium on System on Chip (SoC)*, Tampere, Finland, p. 133-138.

Mahfuz, M.U. and K.M. Ahmed, 2005. A review of micro-nano-scale wireless sensor networks for environmental protection: Prospects and challenges. *Science and Technology of Advanced Materials*, 6(4), p. 302-306.

Maia, G., D.L. Guidoni, A.C. Viana, A.L. Aquino, R.A. Mini, and A.A. Loureiro, 2013. A distributed data storage protocol for heterogeneous wireless sensor networks with mobile sinks. *Ad Hoc Networks Journal*, 11(5), p. 1588-1602.

Mekki, K., W. Derigent, A. Zouinkhi, E. Rondeau, and M.N. Abdelkrim, 2014. Improvement of counter-based broadcasting scheme for long communiucating material lifetime. *International Conference on Automation, Control, Engineering and Computer Science (ACECS'14)*, Monastir, Tunisia.

Mir, Z. H. and Y.B. Ko, 2006. A quadtree-based data dissemination protocol for wireless sensor networks with mobile sinks. *Personal Wireless Communications Journal*, 4217, p. 447-458.

Miranda, H., S. Leggio, L. Rodrigues, and K. Raatikainen, 2006. A power-aware broadcasting algorithm. *17th IEEE International Symposium on Personal, Indoor and Mobile Radio Communications*, Helsinki, Finland, p. 1–5.

Mukhtar, H., B. W. Kim, B.S. Kim, and S.-S. Joo, 2009. An efficient remote code update mechanism for Wireless Sensor Networks. *IEEE Military Communications Conference (MILCOM'2009)*, Boston, USA, p. 1-7.

Neumann, J., N. Hoeller, C. Reinke, and V. Linnemann, 2010. Redundancy infrastructure for service-oriented wireless sensor networks. *9th IEEE International Symposium on Network Computing and Applications*, Cambridge, Massachusetts, USA, p. 269-274.

Ribeiro, N.S., Marcos A.M. Vieira, Luiz F.M. Vieira, and O. Gnawali, 2014. CodeDrip: Data Dissemination Protocol with Network Coding for Wireless Sensor Networks. *Wireless Sensor Networks Lecture Notes in Computer Science*, 8354, p. 34-49.

Rumín, A.C., M.U. Pascual, R.R. Ortega, and D.L. López, 2010. Data centric storage technologies: Analysis and enhancement. *Sensors Journal*, 10(4), p. 3023-3056.

Sanchez, E.R., M. Rebaudengo, and L. Zhang, 2011. Performance evaluation of reliable and unreliable opportunistic flooding in wireless sensor network. *17th IEEE International Conference on Networks*, Singapore, p. 7-12.

Sekkas, O., D. Piguet, C. Anagnostopoulos, D. Kotsakos, G. Alyfantis, C. Kassapoglou-Faist, and S. Hadjiethymiades, 2010. Probabilistic information dissemination for MANETs: the IPAC approach. *20th Tyrrhenian Workshop on Digital Communication*, Pula, Italy, p. 375-385.

Shen, C., S. Harte, E. Popovici, B. O'Flynn, R.C. Atkinson, and A.G. Ruzzelli, 2009. Automated protocol selection for energy efficient WSN applications. *Electronics Letters*, 45(21), p.1098-1099.

Shen, H., L. Zhao, and Z. Li, 2011. A distributed spatial-temporal similarity data storage scheme in wireless sensor networks. *IEEE Transactions on Mobile Computing*, 10, p. 982-996.

Shim, G. and D. Park, 2006. Locators of mobile sinks for wireless sensor networks. *International Conference on Parallel Processing Workshops (ICPPW'06)*, Columbus, USA, p. 159-164.

Sundani, H., H. Li, V. Devabhaktuni, M. Alam, and P. Bhattacharya, 2011. Wireless Sensor Network Simulators, A Survey and Comparisons. *Journal of Computer Networks*, 2(5), p. 249-265.

Vecchio, M., A.C. Viana, A. Ziviani, and R. Friedman, 2010. DEEP: Density based proactive data dissemination protocol for wireless sensor networks with uncontrolled sink mobility. *Journal of Computer Communications*, 33(8), p. 929-939.

Viana, A.C., T. Herault, T. Largillier, S. Peyronnet, and F. Zaïdi, 2010. Supple: A Flexible Probabilistic Data Dissemination Protocol for Wireless Sensor Networks. *13th ACM international conference on Modeling, analysis, and simulation of wireless and mobile systems*, Bodrum, Turkey, p. 385-392.

Zhi-yan, C., J. Zhen-zhou, and H. Ming-zeng, 2007. An energy-aware broadcast scheme for directed diffusion in wireless sensor network. *Journal of Communication and Computer*, 4(5), p. 28-35.